\documentclass[%
reprint,
 amsmath,amssymb,
 aps,
prb,
]{revtex4-2}

\usepackage{siunitx}
\usepackage{makecell}
\usepackage{dcolumn}
\usepackage{graphicx}
\usepackage{bm}
\usepackage[colorlinks,linkcolor=black,anchorcolor=blue, urlcolor=blue,citecolor=blue]{hyperref}

\usepackage{mathptmx}

\usepackage{xspace} 
\newcommand{\GWGamma}{\ensuremath{G_0W_0\Gamma^{(1)}_0}}
\newcommand{\SOSEX}{\ensuremath{G_0W_0+}SOSEX}
\newcommand{\GW}{\ensuremath{G_0W_0}\xspace}
\newcommand{\GWA}{GWA~}
\newcommand{\scGW}{sc\ensuremath{GW}\xspace}

\newcommand{\GWGammas}{\ensuremath{G_0W_0\Gamma^{(1)}_0}\xspace}
\newcommand{\SOSEXs}{\ensuremath{G_0W_0+}SOSEX\xspace}
\newcommand{\GWs}{\ensuremath{G_0W_0}\xspace}

\usepackage{booktabs}   
\usepackage{multirow}   
\usepackage[referable]{threeparttablex}
\newcolumntype{R}{>{\RaggedRight}X} 
\newcolumntype{x}{>{$}l<{$}}  
\newcolumntype{y}{>{$}c<{$}}  
\newcolumntype{z}{>{$}r<{$}}  

\hypersetup{pdftitle={the FSOS-apply draft}}
\begin{document}
\title{Vertex effects in describing the ionization energies of the first-row transition-metal 
monoxide molecules}
\author{Yanyong Wang}
\affiliation{CAS Key Laboratory of Quantum Information, University of Science and Technology of China, Hefei, Anhui 230026, China}
\author{Xinguo Ren}\email{renxg@iphy.ac.cn}
\affiliation{Institute of Physics, Chinese Academy of Sciences, Beijing 100190, China}
\affiliation{Songshan Lake Materials Laboratory, Dongguan 523808, Guangdong, China}

\date{\today}
\begin{abstract}
The $GW$ approximation is considered to be the simplest approximation with Hedin's formulation of many-body perturbation theory.
It is expected that some of the deficiencies of the $GW$ approximation can be overcome by adding the so-called vertex corrections. 
In this work, the recently implemented \GWGammas scheme, which incorporates the vertex effects by adding 
the full second-order self-energy correction to the $GW$ self-energy, is applied to a set of first-row transition-metal monoxide (TMO) anions. 
Benchmark calculations show that results obtained by \GWGammas on top of the B3LYP
hybrid functional starting point (SP) are in good agreement with experiment data, giving a mean absolute error 
of 0.13 eV for a testset comprising the ionization energies (IEs) of 27 outer valence molecular orbitals (MOs) from 9 TMO anions. 
A systematic SP-dependence investigation by varying the ratio of the exact exchange (EXX) component in the PBE0-type SP reveals that, 
for \GWGammas,  the best accuracy is achieved with $20\%$ EXX. Further error analysis in terms of the orbital symmetry characteristics
(i.e, $\sigma$, $\pi$, or $\delta$) in the testset indicate the best amount of EXX in the SP for  \GWGammas calculations
is independent of MO types, and this is in contrast with the situation of \GW calculations where the best EXX ratio varies
for different classes of MOs. 
Despite its success in describing the absolute IE values, we however found that \GWGammas faces difficulties in describing the
energy separations between certain states of interest, worsening the already underestimated \GW predictions.
\end{abstract}
\maketitle

\section{Introduction\label{sec:intro}}
Among various electronic structure methods available in computational chemistry and materials science, the $GW$ approximation (GWA)
~\cite{Hedin1965/PhysRev.139.A796,Aryasetiawan1998/RepProgPhys.61.237} plays a central role in describing excitation energies in
the one-electron attachment or detachment processes. Different from the Hartree-Fock theory, the \GWA further accounts for
the dynamic screening effects at the level of the random phase approximation (RPA), i.e., due to the creation of noninteracting electron-hole pairs 
as a linear response to the propagation of an external charge in an interacting electron cloud~\cite{Reining2018/WIREsComputMolSci.8.e1344}. Exploiting
the screening concept and error cancellations, the perturbative $GW$ approach based on references provided by
(generalized) Kohn-Sham density functional theory ((g)KS-DFT)~\cite{Kohn1964/PhysRev.136.B864,Kohn1965/PhysRev.140.A1133,GDFT:1996} 
has achieved great success in its applications to weakly and moderately correlated systems~\cite{Louie1985/PhysRevLett.55.1418,Louie1986/PhysRevB.34.5390,Godby1988/PhysRevB.37.10159,Onida2002/RevModPhys.74.601,Blase2011/PhysRevB.83.115103,Golze2020/JPhysChemLett.11.1840,Setten2015/JChemTheoryComput.11.5665,Bruneval2013/JChemTheoryComput.9.324,Knight2016/JChemTheoryComput.12.615,Marom2012/PhysRevB.86.245127}. Unsurprisingly, the $GW$ approximation still suffers from a number of limitations~\cite{Reining2009/JChemPhys.131.154111,Reining2018/WIREsComputMolSci.8.e1344,Godby2007/PhysRevA.75.032505,Marom2012/PhysRevB.86.245127}. 
For example, it is known that one-shot $G_0W_0$ calculation had a dependence on the starting point (SP), i.e., the preceding (g)KS references~\cite{Marom2012/PhysRevB.86.245127,Bruneval2013/JChemTheoryComput.9.324,Knight2016/JChemTheoryComput.12.615,Yang2019/JPhysChemLett.10.447}, 
which limits its reliability as a predictive method. This is particularly true for transition metal oxides (TMO): The \GW approach based on
a (semi)local density functional may underestimate or even fail to predict the fundamental band gaps  \cite{Gatti2007/PhysRevLett.99.266402,Carter2014/doi:10.1007/128_2013_503,Bechstedt2009/PhysRevB.79.235114}, while when starting from the DFT$+U$ or hybrid functional references, improvements can be achieved, but the results are still inconclusive~\cite{Bechstedt2009/PhysRevB.79.235114,Jianghong2010/PhysRevB.82.045108,Fuchs2007/PhysRevB.76.115109,Bechstedt2009/PhysRevB.79.235114,Reining2015/PhysRevB.91.045102}. The self-consistent variant of the \GWA, \scGW, though satisfying some conserving laws, could lead to a substantial overestimation of the band gaps and bandwidths for periodic systems, as compared to experimental results~\cite{Kresse2018/PhysRevB.98.155143,Barth1998/PhysRevB.57.2108,Eguiluz1998/PhysRevLett.81.1662}, and in certain cases produces unphysical spectral properties~\cite{Barth1998/PhysRevB.57.2108,Eguiluz1998/PhysRevLett.81.1662,Loos2018/JChemTheoryComput.14.3071}. For molecular systems, \scGW tends to underestimate the ionization energies (IEs)~\cite{Caruso2012/PhysRevB.86.081102,Caruso2013/PhysRevB.88.075105,Koval2014/PhysRevB.89.155417,Caruso2016/JChemTheoryComput.12.5076}, and the deviations increase for larger organic molecules \cite{Caruso2013/PhysRevB.88.075105,Knight2016/JChemTheoryComput.12.615}.

Contributions from higher-order correlation effects, usually called the vertex corrections (VCs), is completely ignored in the \GWA. This is partly
because a direct evaluation of the diagrammatic three-point vertex function following the Hedin's equations 
are rather expensive. Theoretically, VCs are expected to remedy the defects and limitations of the \GWA. In practice, most studies on VCs focus on devising simplified vertices to save computational cost. Nevertheless, VCs derived from a rigorous diagrammatic formulation are still of great interest and
have the advantage that they can be applied to both finite and extended systems~\cite{Ummels1998/PhysRevB.57.11962,Shirley1996/PhysRevB.54.7758,Kresse2014/PhysRevLett.112.096401,Leeuwen2016/PhysRevLett.117.206402,Vlcek2021/JChemPhys.154.121101,Visscher2022/PhysRevB.105.125121} unbiasedly. Technically, VCs can be added to either of the polarizablity~\cite{Reining2005/PhysRevLett.94.186402,Berkelbach2019/JChemTheoryComput.15.2925} and the self-energy~\cite{Ren2015/PhysRevB.92.081104,Kresse2017/JChemTheoryComput.13.4765,Ren2021/JChemTheoryComput.17.5140} or both~\cite{Kutepov2016/PhysRevB.94.155101,Hellgren2018/PhysRevB.98.045117}. Recently, attempts have also been made towards 
incorporating the VCs within the self-consistency loop~\cite{Kresse2007/PhysRevLett.99.246403,Kutepov2016/PhysRevB.94.155101,Ohno2016/PhysRevB.94.121116,Pasquarello2021/PhysRevB.103.L161104,Cunningham2021qsgw}. It has been demonstrated that proper VCs can evidently improve the descriptions 
of band gaps~\cite{Kresse2007/PhysRevLett.99.246403,Kutepov2017/PhysRevB.95.195120,Pasquarello2021/PhysRevB.103.L161104,Cunningham2021qsgw}, IEs, and electron affinities (EA)~\cite{Kresse2014/PhysRevLett.112.096401,Kresse2017/JChemTheoryComput.13.4765,Ren2021/JChemTheoryComput.17.5140}. Despite the impressive improvements achieved by including VCs in the screened Coulomb potential for bulk materials, a recent study~\cite{Berkelbach2019/JChemTheoryComput.15.2925} shows that an EOM-CCSD (equation-of-motion coupled-cluster with single and double excitations) conjectured polarizability that includes a diagrammatically based VC, deteriorates the description of  IEs for a small set of molecules, compared to \GW. Recently, the present authors implemented the lowest-order vertex correction to the self-energy $\Sigma$ in Hedin's equations, corresponding to the full self-order self-energy (FSOS) in terms of $W$ ~\cite{Ren2021/JChemTheoryComput.17.5140}, and benchmarked its performance of small molecules composed of main-group
elements. The resultant \GWGammas scheme, $\Sigma^{G_0W_0\Gamma^{(1)}_0}=\Sigma^{G_0W_0} + \Sigma^{\text{FSOS-}W_0}$, proves to be rather accurate in predicting IPs and EAs for such molecules~\cite{Ren2021/JChemTheoryComput.17.5140}, if PBE and PBE0
functionals are chosen as the starting points. However, the performance of \GWGammas for more complex molecules like 
those containing transition-metal (TM) atoms is still unknown.

In fact, even the performance assessment for $GW$ are still largely
limited to $sp$-electron systems in case of molecules, and only a few studies paid attentions to open-shell~\cite{Ohno2016/PhysRevB.94.121116,Koval2021/NewJPhys.23.093027} or strongly correlated ones~\cite{Botti2014/JChemTheoryComput.10.3934,Hung2017/JChemTheoryComput.13.2135,Ogut2018/JChemPhys.149.064306,Rezaei2021/JChemPhys.154.094307,Williams2020/PhysRevX.10.011041}. Recently, Byun \textit{et al.}~\cite{Ogut2019/JChemPhys.151.134305} reported
that a partial self-consistent $GW$ approach, in which only the eigenvalues in $G$ are updated, can achieve a fairly good agreement with experiment, but their calculations are limited to the early and late $3d$-TMO anions. For $3d$-TMO systems, a number of issues, like multiple close-lying excited spin states arising from the unpaired metal electrons, orbitals with different characteristics, hybridization and localization of orbitals to varying degrees, strong correlations of $3d$-electrons, and multi-configurational nature of the wave function~\cite{Bauschlicher1995/TheoretChimActa.90.189,Merer1989/AnnuRevPhysChem.40.407,Gutsev2000/JPhysChemA.104.5374,Harrison2000/ChemRev.100.679,Andrews2009/ChemRev.109.6765}, all present substantial challenges to  electronic structure calculation methods. For example, debates still exist concerning 
the ground state configurations and the assignments of the photoelectron spectra for some $3d$-TMO anions~\cite{Laisheng1995/JChemPhys.102.8714,Hendrickx2009/JPhysChemA.113.8746,Neumark2015/MolPhys.113.2105,Mavridis2011/JChemPhys.134.234308}. In fact, in the past the ground states and the low-lying excited states of $3d$-TMO clusters have been systematically studied by the high-level multireference methods~\cite{Merer1989/AnnuRevPhysChem.40.407,Bauschlicher1986/JChemPhys.85.5936,Mavridis2011/JChemPhys.134.234308,Mavridis2010/JPhysChemA.114.8536,Pykavy2003/JPhysChemA.107.5566,Bauschlicher1983/JChemPhys.79.2975,Mavridis2010/JPhysChemA.114.9333}, DFT methods~\cite{Gutsev2000/JPhysChemA.104.5374,Jinlong2003/JChemPhys.118.9608,Kulik2010/JChemPhys.133.114103}, and experimental  anion photoelectron spectroscopy (PES)~\cite{Laisheng1998/JPhysChemA.102.9129,Laisheng1997/JChemPhys.107.8221,Laisheng1998/JChemPhys.108.5310,Laisheng1995/JChemPhys.102.8714,Weijun2013/ChemPhysLett.575.12,Laisheng1997/JChemPhys.107.16,Laisheng1997/JPhysChemA.101.2103,Laisheng2001/JPhysChemA.105.5709,Jarrold2001/ChemPhysLett.341.313}. In this work we study the vertex effects beyond the \GWA in describing the charged excitations of the first-row transition-metal monoxide anions, and thereby we extend the assessment 
of the \GWGammas scheme
 to TMO molecules.

The rest of this paper is organized as follows: in Section~\ref{sec:comp_details} the computational settings and 
technical details are described. Then in Section~\ref{sec:result_disscuss}, the \GW and \GWGammas results on the ionization energies of low-lying states for the $3d$-TMO testset will be presented and discussed. Here, special attention will be
paid to the role of exact exchange (EXX) in the SPs, whereby the best amount of EXX before and after the inclusion of VCs in the
computational schemes will be identified. A brief summary of this work will be given in Section~\ref{sec:conclusion}.

\section{Computational Details\label{sec:comp_details}}
The geometries of the $3d$-TMO anions are optimized by KS-DFT within the Perdew-Burke-Ernzerhof (PBE)~\cite{Perdew1996/PhysRevLett.77.3865} generalized gradient approximation (GGA), which proves to be reliable for these systems~\cite{Gutsev2000/JPhysChemA.104.5374,Rezaei2021/JChemPhys.154.094307}. The FHI-aims-2020 \textit{tier}-2 numerical
atomic orbital (NAO) basis sets for metal atoms and \textit{tier}-3 NAO basis set for the oxygen atom are used for geometry optimization calculations. 
Two different hybrid functionals, i.e., PBE0~\cite{Adamo1999/JChemPhys.110.6158} and B3LYP~\cite{Becke1993/JChemPhys.98.5648,Stephens1994/JPhysChem.98.11623} are used as the SPs for \GW and 
\GWGammas calculations. For molecules containing TM atoms, PBE0 is a fairly good choice for \GW calculations~\cite{Botti2014/JChemTheoryComput.10.3934}.
As the role of exact exchange (EXX) is crucial for systems with localized $d$ orbitals, for comparison we will also test the popular B3LYP functional 
(with $20\%$ EXX component), as well as the PBE0 family of functionals with varying 
EXX contributions as the SPs. For all the \GW and \GWGammas calculations, the Dunning's augmented correlation-consistent 
``aug-cc-pV5Z''~\cite{Dunning1989/JChemPhys.90.1007} Gaussian basis sets are used for all elements. Additionally, the modified Gauss-Legendre grid~\cite{Ren2012/NewJPhys.14.053020} with 240 points along the imaginary frequency axis is adopted to evaluate both the self-energy and the screened Coulomb interaction. As the self-energy is initially computed on the imaginary axis, an analytical continuation procedure 
is applied to transform the \GW or \GWGammas self-energy into the real axis using the
 Pad\'e approximant~\cite{Baker1975} with 32 parameters.  All the calculations are performed using the all-electron FHI-aims~\cite{Blum2009/ComputPhysCommun.180.2175,Ren2012/NewJPhys.14.053020} code package. The implementation details of \GWGammas in FHI-aims can be found in Ref.~\cite{Ren2021/JChemTheoryComput.17.5140}. 

Besides \GWGamma, an alternative beyond-$GW$ scheme is the so-called second-order screened exchange (SOSEX) self-energy correction~\cite{Ren2015/PhysRevB.92.081104}. The SOSEX correlation self-energy can be viewed as an antisymmetrization of
the correlation part of the $GW$ self-energy by interchanging the way that the bare $v$ line and the screened $W$ line
are connected in the Feynman diagram representation \cite{Ren2015/PhysRevB.92.081104}. Previous studies show that, 
with the same SPs, \GWGammas and \SOSEXs yield 
similar IEs for main-group molecules~\cite{Ren2021/JChemTheoryComput.17.5140}. As the testset is extended to $3d$-TMO 
molecules where the presence of localized $d$-electrons poses additional challenges, it would be interesting 
to check if the similar performance still persists for these systems.

\section{Results and Discussions\label{sec:result_disscuss}}
In this section, we will present the main results of the present study, aiming at assessing
the performance of \GW and \GWGammas for
TMO molecules. To this end, we first introduce the $3d$-TMO testset adopted in this work.
As will be detailed below, the testset consists of 27 low-lying valence 
IEs from 9 TMO anions, where
reliable experimental data are available. The performance of the \GW and \GWGammas approaches for describing
the IEs of low-lying molecular states of the $3d$-TMO anions will be analyzed through the mean errors (MEs) 
and mean absolute errors (MAEs) with reference to experimental reference values.
Special attention will be paid to the SP dependence of the \GW and \GWGammas calculations. Specifically, we will monitor
how the results change by varying the ratio of the EXX component in the preceding hybrid functional calculations. 
\begin{table*}[!htp]\centering\renewcommand{\arraystretch}{1.24}
\setlength{\tabcolsep}{0pt}
\begin{threeparttable}
\label{tab:orbtyp_testset}
\caption{Low-lying occupied MOs in the $3d$-TMO anion orbital testset (denoted as TMO27), 
whose IEs are used to benchmark the theoretical methods in this work. The experimental reference IEs (in eV) are provided for
the selected MOs of each TMO anion.}
\begin{tabular*}{.7\linewidth}{@{\extracolsep{\fill}}ryyyyyyyyyyy}
\toprule
 system &\#\text{ of orbitals} &\multicolumn{9}{c}{\text{experimental IE (eV)\tnotex{expt_vals}}} \\
\cline{3-12}
  &&9\sigma &4\pi &10\sigma_\alpha &9\sigma_\alpha &9\sigma_\beta &1\delta_\alpha &1\delta_\beta &4\pi_\alpha &4\pi_\beta  &3\pi_\beta \\
  ScO$^-$  &1  &1.35 \\
  TiO$^-$  &3  &&&&1.73  &1.30  &2.00 \\
  VO$^-$   &3  &&&&1.93  &1.23  &2.40 \\
  CrO$^-$  &3  &&&&2.13  &            &2.64  &&1.12 \\
  MnO$^-$  &2  &&&&            &1.38  &&&&          &3.58 \\
  FeO$^-$  &5  &&&&2.36  &1.50  &3.39   &1.98 &2.56 \\
  CoO$^-$  &4  &&&&2.17  &1.54  &&1.45  &2.30 \\
  CuO$^-$  &2  &2.75     &1.78 \\
  ZnO$^-$  &4  &&&2.09  &&3.89  &&&2.71 &2.40 \\
\midrule
\bottomrule
\end{tabular*}
\begin{tablenotes}
\item[a]\label{expt_vals} Experimental IE values are taken from: Ref.~\cite{Laisheng1998/JPhysChemA.102.9129} for ScO$^-$, Ref.~\cite{Laisheng1997/JChemPhys.107.8221} for TiO$^-$, Ref.~\cite{Laisheng1998/JChemPhys.108.5310} for VO$^-$, Ref.~\cite{Laisheng2001/JChemPhys.115.7935} for CrO$^-$, Ref.~\cite{Laisheng2000/JChemPhys.113.1473} for MnO$^-$, Ref.~\cite{Laisheng1996/JAmChemSoc.118.5296} for FeO$^-$, Ref.~\cite{Weijun2013/ChemPhysLett.575.12} for CoO$^-$, Ref.~\cite{Laisheng1997/JPhysChemA.101.2103} for CuO$^-$, and Ref.~\cite{Jarrold2001/ChemPhysLett.341.313} for ZnO$^-$, respectively.
\end{tablenotes}
\end{threeparttable}
\end{table*}

\subsection{The TMO27 testset \label{subsec:testset}}
In Table~\ref{tab:orbtyp_testset}, we assemble a total of 27 frontier MOs, corresponding to selected 
low-lying molecular states of the first-row TMO anions. These energy levels have unambiguous MO assignments, established 
via a comparative analysis of their experimental PES spectra and theoretical calculations. 
The symmetry characteristics of these MOs are explicitly given in Table~\ref{tab:orbtyp_testset}.
For notational convenience, below we shall term this testset as ``TMO27".
Except for ScO$^-$ and CuO$^-$, all other TMO anions have open-shell electronic structures. Specifically, TMO27 
contains 14 $\sigma$-orbitals, 7 $\pi$-orbitals and 6 $\delta$-orbitals, belonging to either up- or down-spin channels.
The experimental IE values for these orbitals, taken from experimental literature \cite{Laisheng1998/JPhysChemA.102.9129,Laisheng1997/JChemPhys.107.8221,Laisheng1998/JChemPhys.108.5310,Laisheng2001/JChemPhys.115.7935,Laisheng2000/JChemPhys.113.1473,Laisheng1996/JAmChemSoc.118.5296,Weijun2013/ChemPhysLett.575.12,Laisheng1997/JPhysChemA.101.2103,Jarrold2001/ChemPhysLett.341.313} and used as the reference data in
the present work, are given in  Table~\ref{tab:orbtyp_testset}.

Although these diatomic TMO anions are small in size and simple in chemical formula, they are in fact an important class of 
systems suitable for benchmarking the performance of theoretical methods.  As mentioned in Sec.~\ref{sec:intro}, it is
the competitions between $3d$- and $4s$-orbitals and the strong correlations among the $3d$ electrons that
give rise to the complexity of the electronic structure of TMO systems. For mid-row TMO systems, 
as the number of $3d$-electrons increases, the amount of electronic spin states grows quickly in a narrow energy window, 
which are manifested in their experimental spectral features. This inevitably increases the  difficulty
in assigning the PES peaks to the theoretically calculated energy levels. In fact, identifying the peaks in PES with
single-electron detachment processes, and associating these with MO energy levels are not entirely trivial.
In the Appendix, we give more details about how these issues are clarified in several prototypical TMO anions such as
 CrO$^-$, MnO$^-$ and CoO$^-$.

\subsection{Performance of \GW, \GWGamma, and \SOSEXs}
We performed \GW, \GWGamma, and \SOSEXs calculations for nine TMO anions using computational setups described in Sec.~\ref{sec:comp_details}. The calculations were done on top of two popular hybrid functional (PBE0 and B3LYP)
starting points.
Due to the large self-interaction error 
in systems containing TM elements~\cite{Marom2011/PhysRevB.84.195143,Ogut2018/JChemPhys.149.064306}, the commonly used PBE functional is not a
proper SP. As a matter of fact, tentative calculations indicate that, for both \GW and \GWGammas calculations with the PBE SP, it's rather hard and often impossible to obtain converged 
quasiparticle (QP) values when solving the QP equation iteratively, due to
the complex structure of the self-energy near the solution.

\subsubsection{FeO$^-$: A case study \label{subsec:FeO-}}
\begin{table}[!ht]\centering\renewcommand{\arraystretch}{1.3}
\setlength{\tabcolsep}{0pt}
\begin{threeparttable}
\caption{Ionization energies (in eV) of low-lying states in FeO$^-$, with its ground state established as $^4\Delta\,\,(9\sigma^2 1\delta^3 4\pi^2)$.
After an electron is kicked out from one the MOs listed in the second row, the system becomes charge neutral and resides in a
corresponding electronic state (final state) given in the first row.}
\label{tab:FeO}
\begin{tabular*}{.99\linewidth}{@{\extracolsep{\fill}}lyyyyyy}
\toprule
Final state &^5\Delta      &^5\Sigma^+    &^3\Delta       &^3\Pi       &^3\Sigma^+  &^5\Phi \\
MO    &9\sigma_\beta &1\delta_\beta &9\sigma_\alpha &4\pi_\alpha &1\delta_\alpha &3\pi_\beta \\
Expt\tnotex{FeO_expt}&1.50          &1.98           &2.36        &2.56   &3.39 \\
\cline{1-7}
PBE0      &-0.20  &0.77   &1.20   &1.26   &2.69   &2.38 \\
&\multicolumn{6}{c}{\text{@PBE0}} \\
\cline{2-7} 
\GW       &1.15   &1.69   &2.41   &2.37   &3.64   &3.61 \\
\SOSEX    &-      &2.36   &2.61   &2.85   &4.01   &3.89 \\
\GWGamma  &1.63   &2.29   &2.65   &2.73   &3.95   &3.96 \\
\cline{1-7}
B3LYP     &-0.33  &0.36   &0.88   &0.96   &2.08   &2.20 \\
&\multicolumn{6}{c}{\text{@B3LYP}} \\
\cline{2-7}
\GW       &1.03   &1.29   &2.08   &2.08   &3.06   &3.40 \\
\SOSEX    &-      &2.01   &2.25   &2.51   &3.41   &3.67 \\
\GWGamma  &1.55   &1.90   &2.33   &2.42   &3.36   &3.76 \\
\midrule
\bottomrule
\end{tabular*}
\begin{tablenotes}
\item[a]\label{FeO_expt} Ref.~\cite{Laisheng1996/JAmChemSoc.118.5296}.
\end{tablenotes}
\end{threeparttable}
\end{table}
A general procedure of our study proceeds as follows: 
(1) Identify the ground-state (many-electron) configuration of the anions; 
(2) analyze the experimental PES 
and interpret the individual peaks in PES in terms of the single-particle detachment processes from the
anions' ground state; and (3) associate the experimental peak positions
with the calculated QP energy levels, characterized by MO symmetries as listed in Table~\ref{tab:orbtyp_testset}.
We have performed such analyses for all nine TMO anions.  Below, using the FeO$^-$ anion
as a showcase, we illustrate how the entire calculation and analysis are carried out. To avoid redundancy, we will not 
describe other anions in detail here,  
but provides in the Appendix additional information for the cases where the analyses are not entirely trivial, 
such as CrO$^-$, MnO$^-$, and CoO$^-$. Note that although multi-configurational characters are expected to be reflected in the anion PES of TMO systems, the primary peaks can be properly described by single particle excitations~\cite{Laisheng1995/JChemPhys.102.8714}, and the ground-state wave function is supposed to be dominated by a single Slater determinant~\cite{Gutsev2000/JPhysChemA.104.5374,Chenji2021/JChemPhys.154.164302,Kulik2010/JChemPhys.133.114103}. When there are controversies in the spectral assignments, we have to make a judicious decision based on our own calculation results.

The first step is to identity the ground-state configuration of FeO$^-$ anion, i.e., how the outer valence MOs are occupied by electrons, which further determines the symmetry and spin moment of the system. In case of FeO$^-$ anion, 
various experiments consistently confirm that the low-spin ($S=3/2$) state $^4\Delta$ ($9\sigma^2 1\delta^3 4\pi^2$) is the ground state~\cite{Lineberger1987/JChemPhys.86.1858,Laisheng1995/JChemPhys.102.8714,Drechsler1997/JChemPhys.107.2284,Neumark2015/MolPhys.113.2105}, while multireference and correlation calculations suggest the high-spin ($S=5/2$) $^6\Sigma^+$ state 
($9\sigma^2 4\pi^2 1\delta^2 10\sigma^1$)~\cite{Hendrickx2009/JPhysChemA.113.8746,Mavridis2011/JChemPhys.134.234308} 
as the ground state. 
DFT studies (including PBE calculations in this work) in fact support the $^4\Delta$ proposal~\cite{Gutsev1999/JPhysChemA.103.5812,Gutsev2000/JPhysChemA.104.5374,Uzunova2008/JChemPhys.128.094307}, 
in agreement with experiment.
In Table~\ref{tab:FeO}, we provide an orbital analysis of the experimental data, based on the assumption that $^4\Delta$ is the ground state of the anion,
and the calculated results using hybrid functionals, \GWs, and beyond-\GWs approaches. After one electron is removed from the valence orbitals
as given in the second row of Table~\ref{tab:FeO}, the system becomes a neutral FeO molecule, 
with an electronic configuration (the final state) characterized by the symmetry symbols presented in the first row.

After the initial ground state of the anion is established, there is still the issue of assigning the spectral features to
the final states with one electron removed, which directly corresponds to the energy ordering of the MOs 
(the second row in Table~\ref{tab:FeO})
The original assignment of the spectral feature is based on relatively rough configuration interaction (CI) calculations in the early days~\cite{Krauss1985/JChemPhys.82.5584,Laisheng1995/JChemPhys.102.8714}, while our proposed assignment chiefly follows the ordering of energy levels from hybrid functional calculations. In fact, the experimental assignment for the first two quintet states ($^5\Delta$ and $^5\Sigma^+$) are fairly solid~\cite{Neumark2015/MolPhys.113.2105,Laisheng1995/JChemPhys.102.8714,Lineberger1977/JChemPhys.66.5054}, and
the difference between our assignment and that of Refs.~\cite{Krauss1985/JChemPhys.82.5584,Laisheng1995/JChemPhys.102.8714} is the ordering of the three triplet states ($^3\Delta,\,^3\Pi$ and $^3\Sigma^+$) in Table~\ref{tab:FeO}. Based on the new spectra assignment, for the five peaks in the PES spectra of FeO$^-$ anion~\cite{Laisheng1996/JAmChemSoc.118.5296}, \GWGamma@B3LYP gives the best MAE with 0.07 eV, followed by \GW@PBE0 which also yields a rather good MAE of 0.23 eV. The high accuracy of the \GWGamma@B3LYP (also \SOSEX@B3LYP) predictions for the spectra is a strong evidence supporting our assignment. 

From Table~\ref{tab:FeO}, we can see that the \GW method is problematic in predicting the ordering of $9\sigma_\alpha$ and $4\pi_\alpha$-orbitals. In contrast, both \SOSEXs and \GWGammas restore the right energy ordering, and produce a separation between these two orbitals that compare well
with the experimental results. Another interesting feature is that
the $1\delta_\alpha$ orbital is rather sensitive to the SPs as compared to other orbitals. Chaning
the SP from B3LYP to PBE0 greatly raises ($\sim0.6$ eV) the calculated IE of $1\delta_\alpha$-electron, bringing it
close to the IE of $3\pi_\beta$-electron and rendering the assignment of $1\delta_\alpha$ and $3\pi_\beta$ complicated with all the calculation results obtained with the PBE0 SP. More calculations in subsection~\ref{subsec:bestEXX} reveal that the positions of $\delta$-orbitals are very sensitive to the EXX amount. While with the B3LYP SP, the $1\delta_\alpha$ and $3\pi_\beta$-orbitals are well separated, with both \SOSEXs and \GWGammas predicting a very accurate IE for the $1\delta_\alpha$-electron.

\subsubsection{Overall performance for TMO27}
\label{sec:performance}
\begin{figure*}[!htp]\centering
\includegraphics[scale=1.]{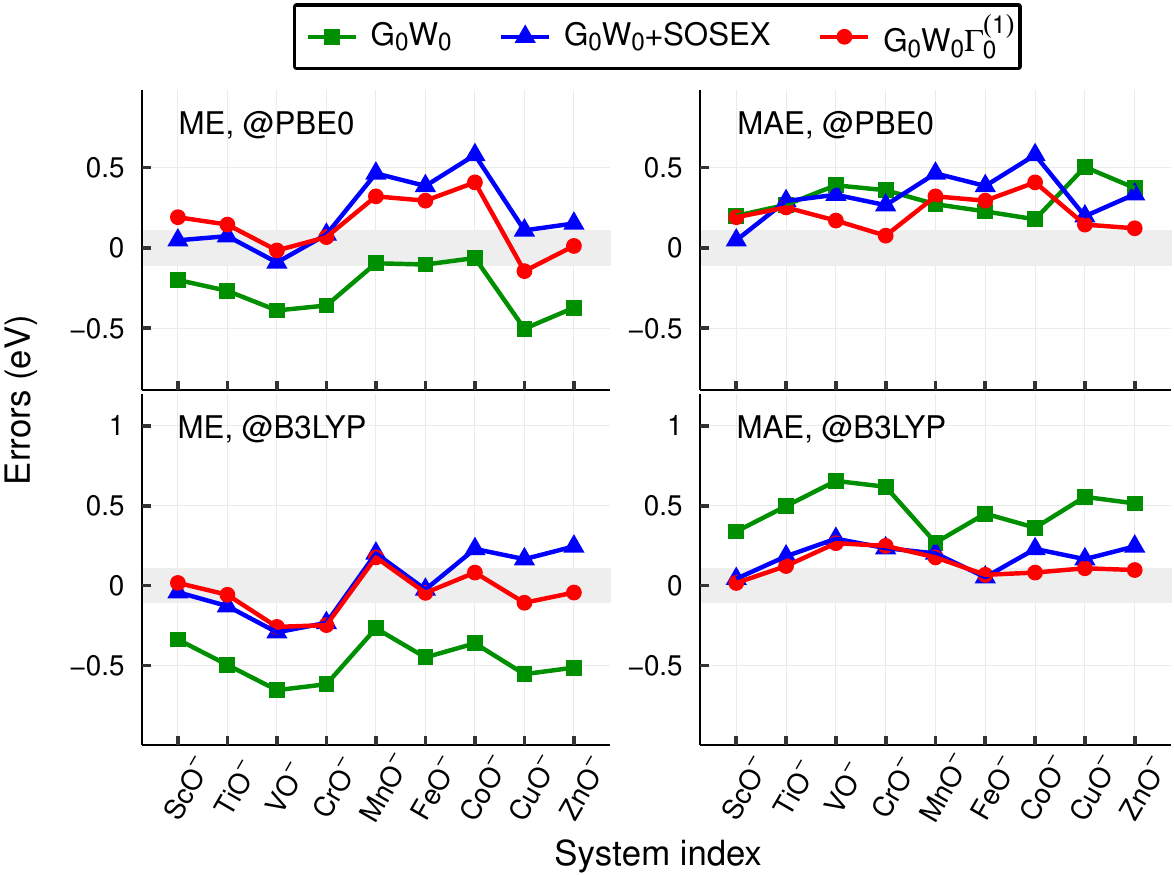}
\caption{The MEs (left panels) and MAEs (right panels) of the \GW, \SOSEX, and \GWGammas results on the TMO27 test set. The correspondent SPs are labelled in each panel. The light grey area refers to the $[-0.1, 0.1]$ eV error window.}
\label{fig:MEsMAEs_per_system}
\end{figure*}

In Fig.~\ref{fig:MEsMAEs_per_system}, we plot the mean errors (ME) and mean absolute errors (MAE) of the calculated
IE results of the low-lying states in the TMO27 testset, with respect to the experimental values for each individual
anion. In fact, for ScO$^{-}$, there is just one MO ($9\sigma$) included, while for others, 
the results are an averaging of the (absolute) errors of 2-5 orbitals included in the testset (cf. Table~\ref{tab:orbtyp_testset}). 
The actual computational and experimental IE values for all the 27 MOs are presented in the supplementary material (SM). Similar to the case of main-group molecules, \GW has a general tendency to underestimate the IEs, although
the magnitude of the underestimation depends on the SPs. It can be seen from Table~~\ref{tab:MEsMAEs} that
\GW@PBE0 shows a better performance than \GW@B3LYP, reducing the MAE from 0.48 eV to 0.30 eV. Figure~\ref{fig:MEsMAEs_per_system} further
reveals that this is because \GW@PBE0 gives higher IEs than \GW@B3LYP, 
bringing the results in better agreement with experiment.

\begin{table}[!htp]\centering\renewcommand{\arraystretch}{1.3}
\setlength{\tabcolsep}{2pt}
\begin{threeparttable}
\caption{ \label{tab:MEsMAEs} The MEs and MAEs (in eV) of the \GW, \SOSEX, and \GWGammas results on the IEs of the TMO27 testset with respect to experimental results. }
\begin{tabular*}{1.\linewidth}{@{\extracolsep{\fill}}lcccccc}
\toprule
SP &\multicolumn{2}{c}{\GW}   &\multicolumn{2}{c}{\SOSEX} &\multicolumn{2}{c}{\GWGamma} \\
\cline{2-3}\cline{4-5}\cline{6-7}
&ME &MAE &\enspace ME &\quad MAE &ME &MAE \\
@PBE0    &$-0.25$ &0.30   &\enspace0.23 &\quad0.36   &\,\,\,0.16 &0.23\\
@B3LYP   &$-0.48$ &0.48   &\enspace0.00 &\quad0.19   &$-0.06$ &0.13 \\
\midrule
\bottomrule
\end{tabular*}
\end{threeparttable}
\end{table}

Upon including the VCs, \GWGammas produces higher IEs than \GW, reducing the deviations between the computational results
and the experimental reference
values, as can be clearly seen  from the left panels of Fig.~\ref{fig:MEsMAEs_per_system}. Roughly speaking, the ME lines of \GWGammas follow the same variation pattern as the \GW ones with the same SPs, suggesting that the
major effect of the VCs is to push the occupied frontier orbitals downwards to higher binding energies. 
The MAE results presented in Fig.~\ref{fig:MEsMAEs_per_system} and Table~\ref{tab:MEsMAEs} also show that,  compared to \GW,  the vertex
corrections in both SOSEX and FSOS schemes systematically improve the agreement with experiment for the description of low-energy excited states. Particularly, \GWGamma@B3LYP gives the lowest MAE of 0.13 eV for the TMO27 testset. Interestingly, both \SOSEXs and \GWGammas calculations perform better on top of B3LYP than PBE0, while the opposite is
true for \GW. Such a general behavior is observed for nearly all individual systems, except for CrO$^-$ which is the only system for
which PBE0 is a better SP than B3LYP for \GWGammas calculations, and for CuO$^-$ and ZnO$^-$ where the PBE0 and B3LYP SPs yield similar results.

Regarding the two VC schemes, the behavior of \SOSEXs generally follows that of \GWGammas, as indicated by the ME and MAE curves in Fig.~\ref{fig:MEsMAEs_per_system}. It should be noted that the \SOSEXs data is absent for $9\sigma_\beta$-electrons, 
due to the problem that the a stable solution cannot be found by iteratively solving the QP equation. Quantitatively, \GWGammas still performs better than \SOSEXs, as can be seen from Table~\ref{tab:MEsMAEs}. Figure~\ref{fig:MEsMAEs_per_system} further reveals that the differences between these two VC schemes are more pronounced with the PBE0 SP than the B3LYP SP, and that the deviations are larger for late TMO anions, with \GWGammas yielding
smaller errors.

\subsubsection{Energy splitting}
\begin{figure}\centering
\includegraphics[scale=1.]{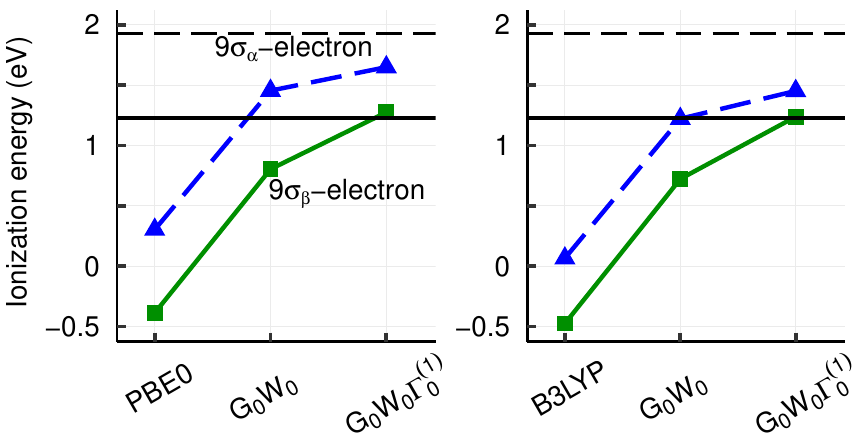}
\caption{The IEs of the two lowest spin orbitals in the VO$^-$ anion as predicted by hybrid functional calculations, \GWs and \GWGammas with corresponding SPs. The dashed and solid lines correspond to the ionization of the $9\sigma_\alpha$ and $9\sigma_\beta$-electrons respectively, while the experimental values for $9\sigma_\alpha$ and $9\sigma_\beta$-electrons are marked by horizontal dashed and solid lines.}
\label{fig:enesplit_VO}
\end{figure}

\begin{table}[!htp]\centering\renewcommand{\arraystretch}{1.5}
\setlength{\tabcolsep}{0pt}
\begin{threeparttable}
\caption{\label{tab:energySep_TiOVO}Energy separations (in eV) of the two lowest spin states (i.e., the splitting between the HOMO and HOMO$-1$ level) in
VO$^-$ and TiO$^-$. Theoretical results are calculated with B3LYP, PBE0, \GW, and \GWGamma, respectively.}
\begin{tabular*}{1.\linewidth}{@{\extracolsep{\fill}}lyyyyyyy}
\toprule
&& &\multicolumn{2}{c}{\text{@B3LYP}} & &\multicolumn{2}{c}{\text{@PBE0}} \\
\cline{4-5} \cline{7-8}
 &\text{Expt\tnotex{nt_expt}} &\text{B3LYP} &\GW &\GWGammas &\text{PBE0} &\GW &\GWGamma \\
VO$^-$   &0.70   &0.54   &0.50   &0.21   &0.69   &0.65   &0.38 \\
TiO$^-$  &0.43   &0.25   &0.23   &0.09   &0.31   &0.28   &0.11 \\
\midrule
\bottomrule
\end{tabular*}
\begin{tablenotes}
\item[a]\label{nt_expt} Experimental data for the TiO$^-$ and VO$^-$ anions are taken from Refs.~\cite{Laisheng1997/JChemPhys.107.8221} 
and \cite{Laisheng1998/JChemPhys.108.5310}, respectively.
\end{tablenotes}
\end{threeparttable}
\end{table}
In the above subsection, the overall performance of \GWs, \SOSEXs, and \GWGammas based on two hybrid functional SPs are discussed. However,
there are important properties which are not reflected in the above overall assessment, namely, 
the energy ordering and spacing between adjacent MOs. 
In Refs.~\cite{Ren2015/PhysRevB.92.081104,Ren2021/JChemTheoryComput.17.5140}, we have pointed out that the description of
the energy separation between the HOMO$-1$ and HOMO$-2$ MOs for the benzene molecule 
is rather challenging for \GW-based methods.
For TMO anions, there has also been considerable interest in the energy spacing between the two lowest spin-states in VO$^-$~\cite{Kulik2010/JChemPhys.133.114103,Shibo2016/PhysRevA.94.062506,MERER1987/JMolSpectrosc.125.465}: the $^4\Sigma^-$ and $^2\Sigma^-$ states, which correspond to the detachments of $9\sigma_\beta$-electron (located at the HOMO level) and $9\sigma_\alpha$-electron (located at the HOMO$-1$ level) from the anion ground state $^3\Sigma$ (with an electronic configuration $9\sigma^2 1\delta^2$), respectively. Table~\ref{tab:energySep_TiOVO} shows that
hybrid functionals, in particular PBE0, produce an energy separation between
$9\sigma_\alpha$ and $9\sigma_\beta$ states that compares fairly well with
the experimental value. When \GWs is applied on top, the energy splitting gets slightly reduced, unfortunately in the wrong direction, compared to
the experimental value. 
Now, after further including the FSOS correction, the energy splitting gets seriously underestimated, 
resulting in a separation that is
only half (in case of PBE0 SP) or even less (B3LYP SP) of the experimental counterpart (cf. the \GWGammas values in Table~\ref{tab:energySep_TiOVO}). 
To have a closer look at what happens, we plot in Fig.~\ref{fig:enesplit_VO} the calculated IEs for both $9\sigma_\alpha$ and
$9\sigma_\beta$ electrons, as determined by the hybrid functionals, \GWs, and \GWGammas. Figure~\ref{fig:enesplit_VO} reveals that  
\GWGammas indeed gives the best absolute IE values compared to experiment, whereas both hybrid functionals and \GWs show appreciable underestimation. 
However, for the latter two types of approaches, the IEs for the two states are underestimated by about the same amount,
and thus the energy separation between the two states come out roughly right. In case of \GWGammas, although the IE of $9\sigma_\beta$ is accurately
predicted, that of $9\sigma_\alpha$ is underestimated by a few tens of meV, resulting in an overall underestimation of the energy separation. 
In fact, similar results with \GWGammas are also found in the case of TiO$^-$ for the energy separation between its HOMO and HOMO$-1$ states. In Table~\ref{tab:energySep_TiOVO}, we list the energy separations given by the calculation methods at different levels for TiO$^-$ and VO$^-$,
which shows that the best prediction is given by PBE0 functional for both systems. When including the VCs, \GWGammas yields energy separations that are only about half of the \GWs results. Varying the SPs does not remedy this problem. This study reveals that although \GWGammas improve over hybrid functionals and \GWs regarding the determination of absolute energy
positions of MOs of TMO anions, it is necessarily so for energy separations between these MOs, in particular between different spin states. 
Finally, we would like to mention that the \SOSEXs
scheme, although performing well for benzene, yields even worse results than \GWGammas for the energy separations for TiO$^-$ and VO$^-$.
This problem needs further studies to achieve a better understanding.
\begin{figure*}[thp]\centering
\includegraphics[scale=1.]{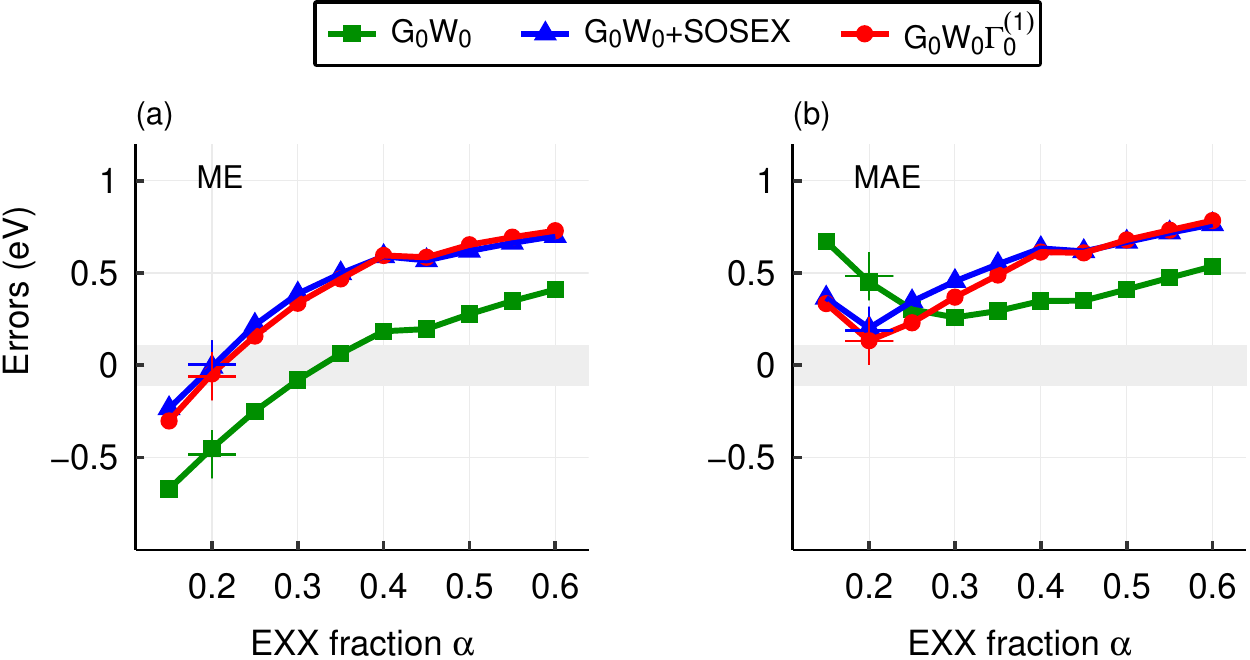}
\caption{MEs (left panel) and MAEs (right panel) of the \GW, \SOSEXs and \GWGammas results on the IEs of low-lying states for the TMO27
testset when varying the EXX ratio from 0.15 to 0.60 in the PBE0-type functional. Additionally, the corresponding \GWs and beyond-\GWs results 
obtained with the B3LYP SP (which has $20\%$ EXX) are labeled with crosses. }
\label{fig:MEsMAEs_varyingEXX}
\end{figure*}

\subsection{Influence of the amount of EXX in the hybrid functional SPs
\label{subsec:bestEXX}}
In this subsection, we devote ourselves to discussing the impact of the amount of EXX in the hybrid functional SPs on the accuracy of the
subsequent \GWs and beyond-\GWs calculations. In the previous section, we have observed that, 
out of the two hybrid functionals, \GWs performs better 
with the PBE0 SP, whereas \SOSEXs and \GWGammas perform better with the B3LYP SP. One of the major differences 
between PBE0 and B3LYP is the ratio of EXX they contain: In PBE0, there is $25\%$ EXX, while in B3LYP, there is $20\%$ EXX. 
To check if it is the different ratio of EXX in these two hybrid 
functionals that caused the difference, we carried out a systematic investigation by varying the ratio of
EXX in the PBE0 functional from $15\%$ to $60\%$, and monitor how the performance changes.
The second motivation of this investigation is to check if the improvement brought by the VCs seen in 
Sec.~\ref{sec:performance} can be achieved by \GWs on top of a better SP.

Specifically, we performed \GW, \SOSEXs and \GWGammas calculations for the TMO27 testset based on the PBE0-type hybrid functionals with the EXX 
ratio $\alpha$ varying from 0.15 to 0.60, each separated by an interval
0.05. The MEs and MAEs of the obtained IE results of the three methods 
with respect to the experimental reference values are plotted in Fig.~\ref{fig:MEsMAEs_varyingEXX} 
as a function of $\alpha$. From the left (ME) panel of Fig.~\ref{fig:MEsMAEs_varyingEXX}, one can see that as $\alpha$
increases, the IE values obtained by all three methods increase steadily, first approaching, and
then exceeding and departing from the reference values. The actual calculated IE values for individual anions 
for all $\alpha$'s 
can be found in the SM. For a given $\alpha$, the \SOSEXs and \GWGammas IE's are always higher than the \GWs counterpart,
and the ME curves of the former are almost a constant upward shift (i.e., roughly independent of $\alpha$)  of the latter.
Comparing the two VC schemes, the ME curves of \SOSEXs and \GWGammas are nearly on top of each other.
One may further notice that there is a kick in the ME curves (and also in the MAE curves in the right panel) at around $\alpha=0.45$.
The underlying reason for this behavior is that the ground state of FeO$^-$ is incorrectly predicted by PBE0 calculations with $\alpha > 0.40$, 
and thereby further PES assignment is meaningless. 
Moreover, in the valid regime of $\alpha$, the FeO$^-$ results are more sensitive to the amount of EXX 
than other systems, hence contributing a significant part to the slope of the ME curves. 
For $\alpha>0.40$, the FeO$^-$ data are excluded, and this results in a noticeable drop of the slope of the ME (MAE) curves.

In the right panel of Fig.~\ref{fig:MEsMAEs_varyingEXX}, we plotted the MAEs of the three schemes 
as a function of $\alpha$. 
The minima of MAE lines signify the best $\alpha$ for each method. The MAE curves show that, for \GWs calculations, the PBE0 with $30\%$ EXX
provides the best SP, whereas for both \SOSEXs and \GWGammas calculations the best fraction of EXX is $20\%$. Furthermore, 
one can see that the minimum of \GWGamma's MAE line is lower than that of \GW's, meaning that with VCs, one
can achieve an accuracy (MAE of 0.13 eV for \GWGamma) that cannot be achieved by \GWs with best tuned SP (MAE of 0.26 eV for \GW). 
For comparison, we also marked in Fig.~\ref{fig:MEsMAEs_varyingEXX} the ME and MAE results obtained with the B3LYP SP at the EXX ratio of 0.2,
where one can see that very similar accuracy is achieved for all three methods on top of the B3LYP SP and on top of 
the PBE0-type functional with 0.2 EXX.
\begin{figure*}\centering
\includegraphics[scale=1.]{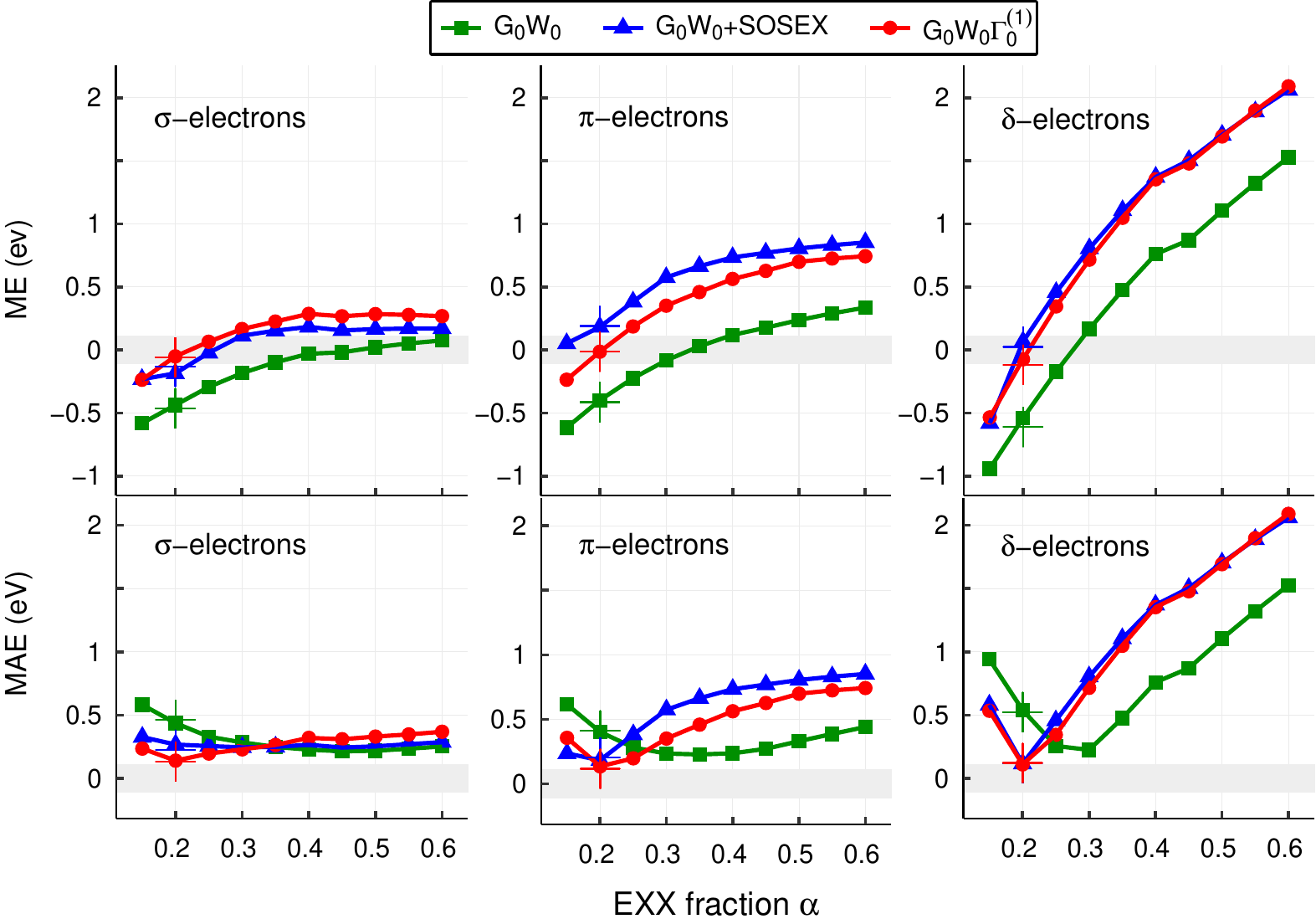}
\caption{MEs (upper panels) and MAEs (lower panels) of the \GW, \SOSEXs and \GWGammas results on the IEs of $\sigma$- (left panels), $\pi$- (middle panels) and $\delta$-orbitals (right panels), respectively, in the TMO27 testset when varying EXX ratios from 0.15 to 0.60 in the PBE0 SP. In addition, the corresponding \GWs and beyond-\GWs results on top of B3LYP are labeled with crosses. }
\label{fig:MEsMAS_orb_varyingEXX}
\end{figure*}

What Fig.~\ref{fig:MEsMAEs_varyingEXX} shows is the overall behavior of the SP dependence of the three methods for the entire TMO27 testset. As described in Sec.~\ref{subsec:testset}, this testset consists of 14 $\sigma$-orbitals, 7 $\pi$-orbitals, and 6 $\delta$-orbitals. It would be interesting
to check if there is any difference in the SP dependence of these three different types of MOs. To this end, in Fig.~\ref{fig:MEsMAS_orb_varyingEXX} 
we plot the ME and MAEs of the three methods separately for the $\sigma$, $\pi$, and $\delta$ types of MOs. 
Firstly, from the ME curves in the upper panels of Fig.~\ref{fig:MEsMAS_orb_varyingEXX}, one can clearly see that the three subsets with different
orbital characters show rather different dependence on the fraction of EXX $\alpha$. 
The degree of sensitiveness grows from $\sigma$ to $\pi$, and to $\delta$.  This indicates that simply adjusting the ratio of EXX may not
be sufficient to reach the best description for all the orbital types, limiting the achievable accuracy 
at the \GW level. Indeed, the MAE lines in the lower panels in Fig.~\ref{fig:MEsMAS_orb_varyingEXX} 
indicate that the best amount of EXX in PBE0-type SP for \GW is 0.45 for $\sigma$-orbitals, 0.35 for $\pi$-orbitals, and 
0.30 for $\delta$-orbitals, respectively. 
Interestingly, for \GWGammas the MAE lines reach minima at a common EXX ratio of 0.20 for all the three MO types, and for \SOSEXs, 
$\alpha=0.20$ is also the best for $\pi$- and $\delta$-orbitals. However, for the $\sigma$ orbitals, the \SOSEXs MAE curve is rather flat and 
does not show a noticeable dependence on
$\alpha$ in a wide range from 0.2 to 0.6. For clarity these results are also summarized in Table~\ref{tab:miniMAEs_EXX_orbtyp}. 
By classifying the results with regard to different orbital types, the distinction between different methods becomes clearer. For example, 
 Fig.~\ref{fig:MEsMAS_orb_varyingEXX} shows
the evolution of the MAEs as a function of the EXX ratio for two VC schemes are rather close for $\delta$-orbitals, while noticeable difference
exists for $\alpha$- and $\pi$-orbitals. Finally, we again marked the B3LYP results in Fig.~\ref{fig:MEsMAS_orb_varyingEXX}, from which one can see similar
accuracy is obtained for all three orbital types for the three methods based on B3LYP and the PBE0 with $20\%$ EXX.
\begin{table}[!htp]\centering\renewcommand{\arraystretch}{1.3}
\setlength{\tabcolsep}{2pt}
\begin{threeparttable}
\caption{ \label{tab:miniMAEs_EXX_orbtyp} The minimal MAEs (in eV) and the corresponding optimal EXX ratios in the PBE0 SP for
the \GW, \SOSEX, and \GWGammas methods in predicting IEs in the three subsets ($\sigma$, $\pi$, and $\delta$) of the TMO27 testset. }
\begin{tabular*}{1.\linewidth}{@{\extracolsep{\fill}}lcycycy}
\toprule
Method &\multicolumn{2}{c}{$\sigma$-electrons} &\multicolumn{2}{c}{$\pi$-electrons} &\multicolumn{2}{c}{$\delta$-electrons} \\
\cline{2-3}\cline{4-5}\cline{6-7}
         &MAE  &\text{EXX}   &MAE &\text{EXX}   &MAE &\text{EXX} \\
\GW      &0.22 &45\%   &0.23 &35\%   &0.23 &30\% \\
\SOSEX   &0.25 &30\%   &0.18 &20\%   &0.12 &20\% \\
\GWGamma &0.14 &20\%   &0.14 &20\%   &0.11 &20\% \\
\midrule
\bottomrule
\end{tabular*}
\end{threeparttable}
\end{table}


\section{Conclusion\label{sec:conclusion}}

In summary, we have performed a detailed study of the vertex effect in beyond-\GWs schemes for describing the outer valence IEs of
$3d$-TMO anions. Using a testset comprising 27 MOs, we demonstrate that including the vertex corrections within schemes like \GWGammas is
advantageous and yields an accuracy that cannot be achieved by simple \GWs schemes. A comprehensive study of the influence of the amount of EXX
contributions in the hybrid functional SPs shows that the \GWGammas scheme performs best at a ratio of $20\%$ for all types of MOs with different
symmetries ($\alpha$, $\pi$, or $\delta$) characteristics, whereas for \GWs, generally a higher fraction is preferred, 
and the best amount varies for different types of MOs.  However, although vertex corrections lead to better absolute IEs for TMO anions, both
\SOSEXs and \GWGammas can deteriorate the description of energy separations between adjacent MOs, like those in VO$^-$ and TiO$^-$, 
compared to \GWs and the hybrid functionals. More investigations are still needed for a better understanding of this issue.

\section*{Data availability}
The data that supports the findings of this study are available within the article and its supplementary material.

\begin{acknowledgments}
This work is supported by National Natural Science Foundation of China (Grant Nos. 12134012, 11874335, 12188101),
 and The Max Planck Partner Group for \textit{Advanced Electronic Structure Methods}.
\end{acknowledgments}

\appendix
\section{Clarification on the one-electron PES assignment}
As mentioned in Sec.~\ref{subsec:testset} and \ref{subsec:FeO-}, the assignment of the experimental PES spectral features to one-electron removal 
energies requires special attentions. Below we provide more information about this aspect for CrO$^-$, MnO$^-$, and CoO$^-$.

\subsection{CrO$^-$}
A key step in the spectral assignment is to identify the ground-state (and in certain cases relevant metastable-state) electronic configurations
from which the electrons are removed.
For CrO$^-$, it has been observed that, in the anion PES experiments, the plentiful spectral features is a blending of photoelectrons emitted from its anion ground state and a metastable anion state~\cite{Laisheng2001/JChemPhys.115.7935,Laisheng2000/JChemPhys.113.1473}. Our PBE calculation 
indicates that the high-spin state $^6\Sigma^+$ with the electronic configuration $9\sigma^1 4\pi^2 1\delta^2$
as the ground state of CrO$^-$, consistent with previous GGA studies~\cite{Gutsev2000/JPhysChemA.104.5374}. Additional PBE0 calculation 
also support this result, with a minor difference of 0.0013 \AA\xspace on the bond length obtained by the two functionals. Further PBE calculations suggest that,
the low-spin $^4\Pi$ state of CrO$^-$ is 0.125 eV higher than $^6\Sigma^+$ in total energy, agreeing well with the experimental separation of 0.096 eV~\cite{Laisheng2001/JChemPhys.115.7935}. As mentioned above, the identification of the anion ground state is important for the subsequent QP calculations, as the spectral assignment is established between the PES peaks and the QP energy levels of the anion ground state, which ultimately determines the performance of the \GWs method and VC schemes.
After the initial state is pinpointed, the energy ordering of individual single-particle orbitals still needs to be sorted out. 
In this regard, two different assignments have been proposed CrO$^-$: one is based on $\Delta$-SCF calculations with GGA functionals~\cite{Laisheng2001/JChemPhys.115.7935}, and the other is the high-level complete-active-space self-consistent-field/multireference configuration interaction (CASSCF/MRCI) calculations~\cite{Gutsev2002/JChemPhys.116.3659}. The orbital ordering predicted by our hybrid functional calculations 
is in line with the CASSCF/MRCI results~\cite{Gutsev2002/JChemPhys.116.3659}, and therefore we adopt their assignment. One of the essential difference between these two calculation results is the relative positions of $9\sigma_\alpha$ and $1\delta_\alpha$ orbitals: GGA predicts the $1\delta_\alpha$ orbital lies higher than $9\sigma_\alpha$~\cite{Laisheng2001/JChemPhys.115.7935} (agreeing with our own PBE calculation) while the order is reversed in MRCI calculations~\cite{Gutsev2002/JChemPhys.116.3659} (agreeing with our hybrid functional calculation results). In fact, the $1\delta_\alpha$-orbital always lies lower than $9\sigma_\alpha$ orbital in the other $3d$-TMO anions with unambiguous assignments of their anion PES spectra. On the other hand, we believe that EXX is crucial in predicting the correct ordering of the frontier orbitals in CrO$^-$, as five parallel-spin electrons in the $^6\Sigma^+$ state could lead to larger exchange interactions.

\subsection{MnO$^-$}
In the original anion PES experimental work of MnO$^-$~\cite{Laisheng2000/JChemPhys.113.1473}, four low-lying states which are relevant to the detachments of $9\sigma_\beta,\,4\pi_\alpha,\,9\sigma_\alpha$ and $3\pi_\beta$-electrons are identified in its spectral assignment with the $\Delta$-SCF calculations for the anion ground state $^5\Sigma^+$. However, we only include $9\sigma_\beta$ and $3\pi_\beta$-orbitals into the TMO27 testset for the sake of reliability, whereas $9\sigma_\alpha$ and $4\pi_\alpha$-orbitals are excluded. This is because the relative ordering of $4\pi_\alpha$ and $9\sigma_\alpha$-orbitals is rather controversial in the theoretical calculations: At the mean-field level, both PBE0 and B3LYP produce almost the same IEs for $4\pi_\alpha$ and $9\sigma_\alpha$-electrons. By contrast, \GW gives a 0.10 eV separation with the $9\sigma_\alpha$-orbital lying below in energy,
whereas \GWGammas and \SOSEXs predict a $0.05\sim0.3$ eV splitting with a reversed ordering. This makes the assignment of the IEs of $4\pi_\alpha$ and $9\sigma_\alpha$-electrons highly 
unreliable. On the other hand, spectral features at 1.38 eV and 3.58 eV, each with large separation from its adjacent peaks, can be
unambiguously assigned to the detachments of $9\sigma_\beta$ and $3\pi_\beta$-electrons, supported by the calculation results.

\subsection{CoO$^-$\label{appendix:CoO-}}
There is a controversy about the ground state of CoO$^-$, i.e., whether it's the low-spin state $^3\Sigma^-\enspace(9\sigma^2 1\delta^4 4\pi^2)$~\cite{Gutsev2000/JPhysChemA.104.5374} or the high-spin state $^5\Delta\enspace(10\sigma^1 9\sigma^2 1\delta^3 4\pi^2)$~\cite{Uzunova2008/JChemPhys.128.094307,Mavridis2012/JPhysChemA.116.6935}. Our own spin-unrestricted 
structural relaxation with the PBE functional indicates that the $^3\Sigma^-$ state as the ground state of CoO$^-$. 
In Refs.~\cite{Weijun2011/JChemPhys.135.134307,Weijun2013/ChemPhysLett.575.12}, the anion PES data of CoO$^-$ was reported, 
but no detailed assignment is provided. Here we provide a 
spectral assignment based on their experimental results and our own calculation work (see Table~\ref{tab:CoO}). It should be mentioned that, Zheng \textit{et al.}~\cite{Weijun2013/ChemPhysLett.575.12} assign the first strong X peak at $1.54\pm0.04$ eV as the EA of CrO, which is quite different from previous two PES experimental work~\cite{Adam1987/JChemPhys.86.5231,Laisheng1999/JChemPhys.111.8389} who recognize the peak at $1.45\pm0.01$ eV as the EA. Instead, Zheng \textit{et al.} regard the peak at 1.45 eV (which has been observed in the 532 nm spectrum in their work) as a hot band, while this feature in fact becomes
as strong as the X peak in the 266 nm spectrum~\cite{Weijun2013/ChemPhysLett.575.12}. Combining the calculation results in this work, we conclude that the peak at 1.45 eV corresponds to the one-electron detachment of the $1\delta_\beta$-electron from the anion ground state $^3\Sigma^-$, and the final state $^4\Delta$ is exactly the ground state of the neutral CoO.
\begin{table}[htp]\centering\renewcommand{\arraystretch}{1.24}
\setlength{\tabcolsep}{0pt}
\begin{threeparttable}
\caption{Ionization energies (in eV) of low-lying states in CoO$^-$ with the anion ground state established as $^3\Sigma^-\,\,(9\sigma^2 1\delta^4 4\pi^2)$.}
\label{tab:CoO}
\begin{tabular*}{0.8\linewidth}{@{\extracolsep{\fill}}lyyyy}
\toprule
\midrule
Final state &^4\Delta        &^4\Sigma^-      &^2\Sigma^+       &^2\Pi \\
MO    &1\delta_{\beta} &9\sigma_{\beta} &9\sigma_{\alpha} &4\pi_{\alpha} \\
Expt\tnotex{CoO_expt} &1.45   &1.54   &2.17   &2.30 \\
\midrule
PBE0      &0.37    &-0.12  &1.22   &1.22 \\
&\multicolumn{4}{c}{@PBE0} \\
\cline{2-5}
\GW        &1.23   &1.30   &2.37   &2.33 \\
\SOSEX     &2.21   &-      &2.66   &2.79 \\
\GWGamma   &1.95   &1.80   &2.67   &2.69 \\
\cline{1-5}
B3LYP     &-0.08   &-0.27  &0.87   &0.93 \\
&\multicolumn{4}{c}{@B3LYP} \\
\cline{2-5}
\GW        &0.84   &1.15   &2.01   &2.03 \\
\SOSEX     &1.89   &-      &2.26   &2.46 \\
\GWGamma   &1.50   &1.68   &2.27   &2.36 \\
\bottomrule
\end{tabular*}
\begin{tablenotes}
\item[a]\label{CoO_expt}Ref.~\cite{Weijun2013/ChemPhysLett.575.12}.
\end{tablenotes}
\end{threeparttable}
\end{table}


\bibliography{TMO_FSOS}
\end{document}